# Universal shape profiles of earthquake ruptures

Amit P. Mehta, Karin A. Dahmen, and Yehuda Ben-Zion[*]

*Department of Physics, University of Illinois at Urbana-Champaign, 1110 West Green Street, Urbana, IL 61801-3080*

*Department of Earth Sciences, University of Southern CA, Los Angeles, CA 90089-0740*

**Universal shape profiles in a variety of systems contain crucial information on the underlying dynamics. We develop such shape profiles for earthquakes as a stronger test of theory against observations. The earthquake analysis shows good agreement between theory and observations, with a discrepancy in one universal profile, just as in magnets. The analysis provides a stronger support for the idea that earthquakes are associated with criticality rather than event statistics only. The results point to the existence of deep connections between the physics of avalanches in different systems.**

*Introduction:* Earthquake phenomenology is characterized by several power law distributions. For example, the frequency-moment distribution $n(M_0)$ of regional and global earthquakes [1] has the form:[2,3]

$$n(M_0) \sim M_0^{-1-b} \tag{1}$$

where the seismic moment $M_0 \sim \sum_i \Delta u_i \Delta A_i$ with $\Delta u_i$ and $\Delta A_i$ being the local slip and rupture area during an earthquake, respectively, and the sum extends over the entire fault. In terms of the magnitude $M$ with $M \sim 2/3\log(M_0)$ for large earthquakes[4], eqn. (1) is rewritten as:

$$\log(n(M)) = a - bM \tag{2}$$

where the constant $a$ characterizes the overall rate of activity in a region and the "$b$-value" $b = 1.5\overline{b}$ gives the relative rates of events in different magnitude ranges. Observed $b$-values of global earthquakes with depth $\leq 50$ km and horizontal-shear (strike-slip), compressional and extensional mechanisms are about 0.75, 0.85, and 1.05, respectively[5].

Another observed power law is the modified Omori law for aftershock decay rates around large rupture zones[2,3]:

$$\Delta N / \Delta t \sim K(t+c)^{\overline{p}} \tag{3}$$

where $N$ is the cumulative number of aftershocks, $t$ is the time after the mainshock, and $K, c,$ and $p$ are empirical constants. In observations[6,7] $p$ is roughly 1 ($0.9 \leq p \leq 1.3$). The association of earthquake statistics with power law relations like Eq. (1) and Eq. (3) led some to suggest that earthquake dynamics is associated with an underlying critical point[8-14]. However, power law distributions can be generated by many other mechanisms[15,16] and it is important to develop criteria that can provide stronger evidence for or against the association of earthquakes with criticality.

Recently enough data have been collected to extract statistics of earthquakes on *individual* fault zones occupying long (order 100 km) and narrow (order 10 km) regions of space. Wesnousky and collaborators[9,10] found that the frequency-size statistics of earthquakes on highly irregular fault zones, with many offsets and branches, as the San Jacinto fault in California, are also described by the Gutenberg-Richter power law relation up to the largest events. However, relatively regular fault zones (presumably generated progressively with increasing accumulated slip over time), such as the San Andreas fault in California, display power-law frequency-size statistics only for small events. These occur in the time intervals between roughly quasi-periodic earthquakes of a much larger ``characteristic'' size that is related to large-scale dimensions of the fault zone [2,9,10,17-19]. Earthquakes of intermediate magnitude are typically not observed on these faults (other than, perhaps, during aftershock sequences). The corresponding frequency size statistics are called the "characteristic earthquake" distribution.

Previously these two types of behaviour on individual fault zones have been modelled as statistics close-to and far-from an underlying critical point[8,20], using a model for a strike-slip fault (FIG. 1) that incorporates long-range interaction and strong heterogeneities[17-19]. The different dynamic regimes were associated with competition between failure-promoting effects of elastic stress-transfer or dynamic weakening, and the opposing effect of strength inhomogeneities in the fault structure. Fisher et al.[8] found that near the critical point the frequency-size statistics follow a power law distribution (with a cutoff at large magnitude), with the same scaling exponent of observed data for strike-slip faults (i.e., a $b$-value of $0.75$). A similar form of frequency-size statistics and predicted $b$-value were obtained also for a critical parameter value in a stochastic branching model[21].

To provide an improved understanding of earthquake dynamics that can suggest additional observables, we focus on faults with Gutenberg-Richter like earthquake statistics (i.e. near-critical behaviour) and develop two universal scaling functions associated with mean moment-rate time profiles at either fixed total moment or fixed total earthquake duration. Universal scaling functions (or shape profiles) give important information on the underlying dynamics, and can be found, for example, in Barkhausen noise in magnets[22], solar flares in astrophysics[23], price fluctuations in financial markets[24], and, as shown here, also in earthquake phenomenology. If the behaviour of fault zones with earthquakes following the Gutenberg-Richter statistics is indeed critical, then the shapes of these functions should be as universal as the exponent $b = 1.5$ in Eq. (1). The scaling functions can be used for a much stronger test of theory versus observed data via comparisons of entire functions as opposed to single-valued exponents. In the following sections we compute the scaling functions for both model predictions and observational data and compare the results.

*The model:* The model we use was developed originally by Ben-Zion and Rice[17-19]: a narrow irregular strike-slip fault zone of horizontal length $L$ and vertical depth $W$ is represented by an array of $N \sim LW$ cells in a two dimensional plane, with constitutive parameters that vary from cell to cell to model the disorder (offsets etc.) of the fault zone structure (FIG. 1). The cells represent brittle patches on the interface between two tectonic blocks that move with slow transverse velocity $v$ in the $x$ direction at a great distance from the fault. The interaction between cells during slip events is governed by 3-D elasticity and falls with distance $r$ from the failure zone as $1/r^3$. These interactions are sufficiently long range that scaling in mean field theory (where the interaction range is set to infinity) becomes exact, up to logarithmic corrections, in the physical fault dimension $(d = 2)$[8,17-19]. In mean field theory, the local stress $t_i$ on a given cell $i$ $(i = 1,\ldots, N)$ is[20]:

$$t_i = J/N \sum_j (u_j - u_i) + K_L(vt - u_i) = J\bar{u} + K_L vt - (K_L + J)u_i \qquad (4)$$

where $u_i$ is the total offset of cell $i$ in the horizontal $x$ direction, $\bar{u} = \sum_{j=1}^{N} u_j / N$ is the average displacement, $J/N$ is the mean-field elastic coupling strength between cells, and $K_L \sim 1/\sqrt{N}$ is the loading stiffness[20] of the moving blocks. Initially the stresses $t_i$ are randomly distributed with $t_{a,i} \leq t_i \leq t_{s,i}$, where $t_{s,i}$ is a fixed local *static* failure threshold stress and $t_{a,i}$ is the fixed local arrest stress. The distributions of static failure stresses and arrest stresses represent the disorder in the fault system. The differences between the failure and arrest stresses give the local distribution of stress drops during brittle failures; the earthquake dynamics depend only on the stress drop distribution. In addition, the scaling behaviour of the system is not sensitive to the exact form of the distributions as long as they are bounded and $t_{a,i} < t_{s,i}$. For additional details see [17-20].

The fault is stuck while the stress on each cell is increased uniformly as $dt_i/dt = K_L v$ as a result of the external loading which is increased adiabatically (that is, we take the limit $v \to 0$). When the stress on a cell reaches its failure threshold $t_{s,i}$, the cell slips by the amount $\Delta u_i = (t_{s,i} - t_{a,i})/(K_L + J)$. This stress drop is uniformly redistributed to all other cells; the resulting stress increase on the other cells can cause some of them to slip as well, leading to an avalanche of cell slips, or an earthquake.

*Phase Diagram*: To include dynamic weakening effects during the failure process[17-19], after an initial slip in an earthquake, the strength of a failed cell is reduced to a *dynamical* value $t_{d,i} \equiv t_{s,i} - e(t_{s,i} - t_{a,i})$, with $0 \leq e \leq 1$ parameterizing the relative importance of dynamical weakening effects in the system. This weakening represents the transition from static to dynamic friction during the rupture and the strength of a failed cell remains at the dynamic value throughout the remainder of the earthquake. In the time intervals between earthquakes all failure thresholds heal back to their static value $t_{s,i}$. At exactly $\varepsilon = 0$ the model produces[8] a power law distribution of earthquake moments $M_0$ following equation (1), cutoff by the finite fault size, with an analytical exponent $\beta = 1/2$ (FIG. 2). This corresponds to a *b*-value of 0.75, close to that associated with observed earthquakes on strike-slip faults[5]. In contrast, for a finite weakening $\varepsilon > 0$ the model produces the characteristic earthquake distribution, with power law statistics for the small events up to a cutoff moment that scales like $M_0^{cutoff} \sim 1/e^2$, and quasi-periodically recurring large characteristic events that scale with the fault size ($M_0 \sim N^{3/2}$).

The model can be expanded further to include dynamic strengthening represented by $\varepsilon < 0$. Ben-Zion and Sammis[25] summarized multidisciplinary observations which indicate that brittle failure of rock has an initial transient phase associated with strain hardening, distributed deformation, and continual creation of new structures. Thus $\varepsilon < 0$ corresponds physically to regions off the main fault segments that are in an early deformation stage. To capture basic aspects of brittle deformation on such regions in the three-dimensional volume around the main fault (FIG. 1), we change the model as follows: when any cell $i$ slips during an earthquake and reduces its stress by $\Delta t_i \equiv t_{f,i} - t_{a,i}$, the failure stress $t_{f,j}$ of *every* cell $j=1,...,N$ is *strengthened* by an amount $|e|\Delta t_i / N$. Once the earthquake is complete the failure stress of each cell is slowly lowered back to its original value. This represents in a simple way the brittle deformation that occurs during an earthquake in the off-fault regions, which are first in a strengthening regime and then have a weakening process. The events that are triggered as the failure stresses are lowered in the weakening period are referred to as *aftershocks*. The occurrence of aftershocks in this version of the model for off-fault regions is in agreement with the observation that a large fraction of observed aftershocks typically occur in off fault regions[26]. For this version of the model with $\varepsilon < 0$, both the primary earthquakes (i.e., mainshocks) and the triggered aftershocks are distributed according to the Gutenberg-Richter distribution, up to a cutoff moment scaling as $1/\varepsilon^2$ (FIG. 2). Assuming that the increased failure stress thresholds $t_{f,i}$ are slowly lowered with time as $\log(t)$ towards their earlier static values $t_{s,i}$, and that the stresses are distributed over a wide range of values, one can show analytically that the temporal decay of aftershock rates at long times is proportional to $1/t$, as in the modified Omori law (3) with $p=1$.

Remarkably, the long length scale behavior resulting from Eq. (4) and the phase diagram of FIG. 2 for $e \leq 0$ can be shown to be the same as that of the mean field version of a driven magnetic domain wall model with infinite range antiferromagnetic interactions[27,28] (the 2 systems are in the same universality class). Similarly in a related model for Barkhausen Noise in hard magnets a very similar phase diagram is seen with high disorder corresponding to $e > 0$ and low disorder corresponding to $e < 0$[29]. The fact that the discussed simple model can capture many of the essential general features of earthquake statistics (or other systems with avalanches, such as driven magnetic domain walls) can be understood through the renormalization group[30,31]. When a model correctly captures certain basic features, such as symmetries, dimensions, and range of interaction, the results provide proper predictions for statistics, critical exponents, and universal scaling functions near a critical point. Consequently, many models that are in the same universality class lead to the same statistics and exponents[8,20,30,31]. The universal scaling functions around the critical point, discussed in the next section, provide additional information that can be used to distinguish between different models and universality classes.

*Universal Mean Moment Rate Time Profiles*: We now focus on fault zones with Gutenberg-Richter power law statistics. Recent analysis allowed researchers to obtain the moment rate $dm_0(t)/dt$, which gives the slip on a fault per unit time during the propagation of earthquake rupture, for hundreds of large seismic events recorded on global networks[32]. Motivated by works on statistical physics of magnetic systems[22,29], we are interested in studying the event-averaged moment rate time profile (FIG. 4) for earthquakes with given duration *T*, denoted with $<dm_0(t|T)/dt>$, and the event-averaged moment rate time profile (FIG. 3) for earthquakes with given total moment $M_0$, denoted with $<dm_0(t|M_0)/dt>$. Here $m_0(t|T)$ is the (cumulative) moment at time $t$ of the propagating earthquake of total duration $T$, and $m_0(t|M_0)$ is the cumulative moment at time $t$ of the earthquake of total moment $M_0$. Theoretical analysis of phase diagrams similar to that shown in FIG. 2 implies that near the critical point there should be, in addition to scaling exponents, also universal scaling functions (up to a rescaling of the ordinate and abscissa)[33]. In our model the two scalable functions in which we are interested, $<dm_0(t|M_0)/dt>$ and $<dm_0(t|T)/dt>$, obey respectively the following scaling relations[33,34]:

$$<dm_0(t|M_0)/dt>/M_0^{1/2} \sim f(t/M_0^{1/2}) \tag{6}$$

and

$$<dm_0(t|T)/dt>/T \sim g(t/T) \tag{7}$$

We determined these scaling functions from corresponding results for magnets, using the fact that the mean field zero-temperature random field Ising model, which is a model for domain wall motion in magnets, is in the same universality class (i.e. has the same universal behaviour on long length scales) as our mean field version of the Ben-Zion and Rice model of Eq. (4) [8,20].

*Exponents and Data Collapses*: We compare the observation results with our model and find remarkable agreement in most cases. The frequency-moment distribution, $D(M_0) \sim M_0^{-1-\beta}$ of the observed data [32] has (inset of FIG. 2) three decades of scaling and an exponent of $\mathbf{b} = 1/2 \pm 0.05$, in close agreement with the model near $\mathbf{e} = 0$. The deviation from power law distribution at the low moment range is associated with the reduced resolution of the observational network for small events. In mean field theory the universal scaling function $f(x)$ in Eq. (6) is of the exact form[34]: $f_{mf}(x) = Axe^{-Bx^2/2}$ with non-universal constants $A=B=1$. In FIG. 3 we present a collapse of the observational data of $<dm_0(t|M_0)/dt>$ for four different values of $M_0$ to obtain the corresponding function $f_{\exp}(x)$ for observations with $x = t/M_0^{1/2}$. We fit the functional form $f_{mf}(x)$ with $A = 4$ and $B = 4.9$ to the collapse of the observed data. Although the observational scaling function $f_{exp}(x)$ deviates from the mean field theory for small values of the ordinate, we find that the mean field exponent 1/2 in the scaling variable $x$ is in close agreement with the observations.

In mean field theory, the function $g(x)$ of Eq. (7) is of the symmetric form: $g_{mf}(x) = Ax(1-x)$, where $A$ is a non-universal constant. In FIG. 4 we collapse observational data for $<dm_0(t|T)/dt>$ for three values of $T$ to obtain the function $g_{exp}(x)$ with $x = t/T$. We also plot the mean field scaling function $g_{mf}(x)$ with $A = 80$. The results show that while the scaling exponents agree, there are notable differences between the observational function $g_{exp}(x)$ and the mean field function $g_{mf}(x)$, especially for small values of the ordinate. The asymmetry in the observed data may result from a rupture process that begins with a failure of a large asperity. This is compatible with observations that hypocenter locations tend to be located close to the area on a fault that produce the largest moment release [e.g., ref. 35]. Additional observational work with a significantly larger data set (not presently available) and corrections to mean field predictions are required to verify the moment rate shape asymmetry and clarify its origin.

Recently it has been shown that the corresponding magnetic domain-wall model[22,29] predicts well the critical scaling exponents for Barkhausen noise experiments in magnets. Significantly, the experimental scaling function for magnetization avalanches or Barkhausen "pulses"[22,29], that is the analogue of the moment-rate time profile for fixed earthquake duration of Eq. (7), shows the same type of asymmetry that is observed for earthquakes (FIG. 4). This raises the possibility that the origin of this asymmetry may be similar in both magnets and earthquakes. Our study shows that there are important theoretical and observational connections between processes in earthquake and magnetic systems. Additional high-resolution observations of small events and theoretical analysis of modifications in the model may provide a deeper understanding of the dynamics of earthquakes and avalanches in other systems of condensed matter physics.

**Acknowledgements**

We are grateful to Susan Bilek for giving us the observational data. We thank James P. Sethna and Michael B. Weissman for very helpful discussions and MBW for first drawing our attention to the possible similarity between the asymmetric universal scaling function in magnets and earthquakes. A.M. and K.D. acknowledge support from NSF grants No. DMR 99-76550 (Materials Computation Center), and No. DMR 0314279, the Alfred P. Sloan foundation (K.D.), and a generous equipment award from IBM. YBZ acknowledges support from a Mercator fellowship of the German Research Society (DFG).

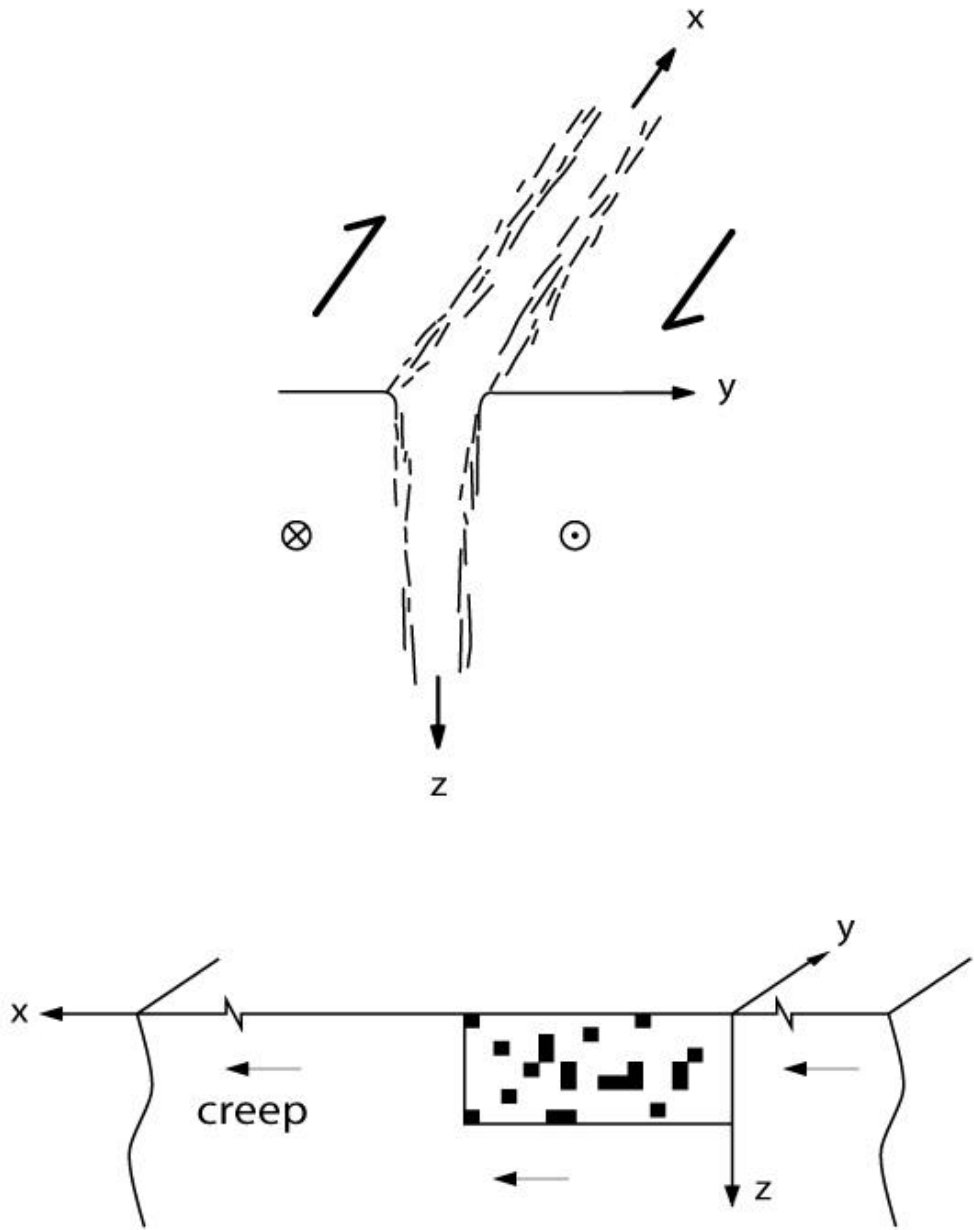

FIGURE 1. A planar representation of a 3-D segmented fault zone by a 2-D heterogeneous fault embedded in a 3-D solid[17-19]. The geometric disorder is modelled as disorder in strength properties of the planar fault.

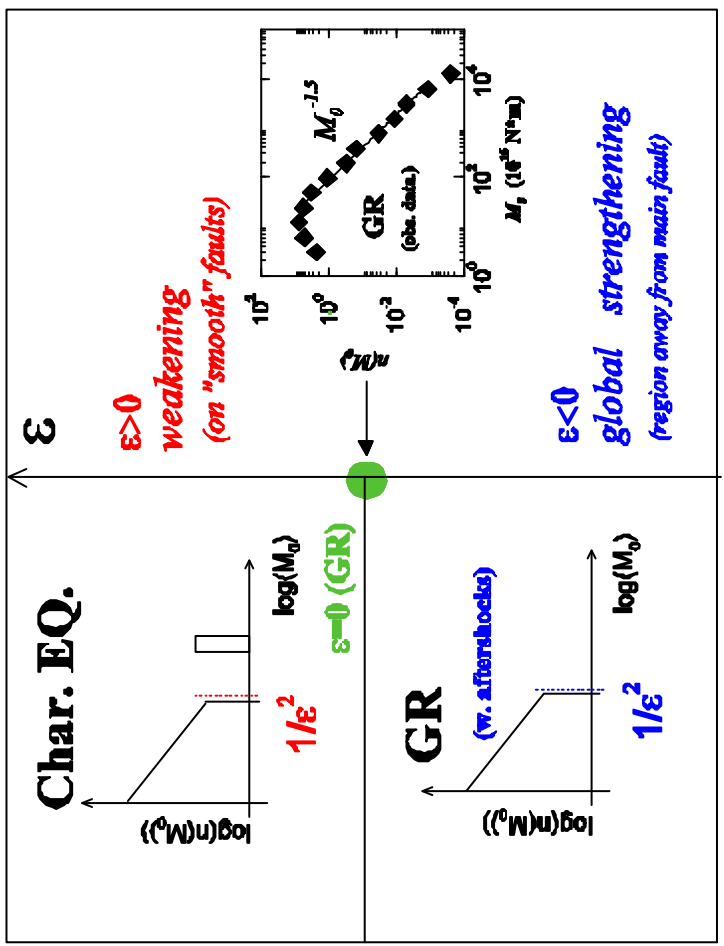

FIGURE 2. Phase diagram of model. The range $\varepsilon > 0$ corresponds to ``mature'' (smooth) localized faults with a weakening rheology and characteristic earthquake statistics. The value $\varepsilon = 0$ corresponds to ``immature'' strongly inhomogeneous fault zones with Gutenberg-Richter statistics. The range $\varepsilon<0$ corresponds to the fracture network away from the main fault, characterised by strengthening due to the creation of new structures and associated emerging aftershocks. Inset: Frequency-moment statistics of the observed earthquakes[32]. The bold line is a power law with an exponent of $-1.5$, which corresponds to the scaling behaviour near $\varepsilon=0$, and also to the avalanche size distribution in mean field theory.

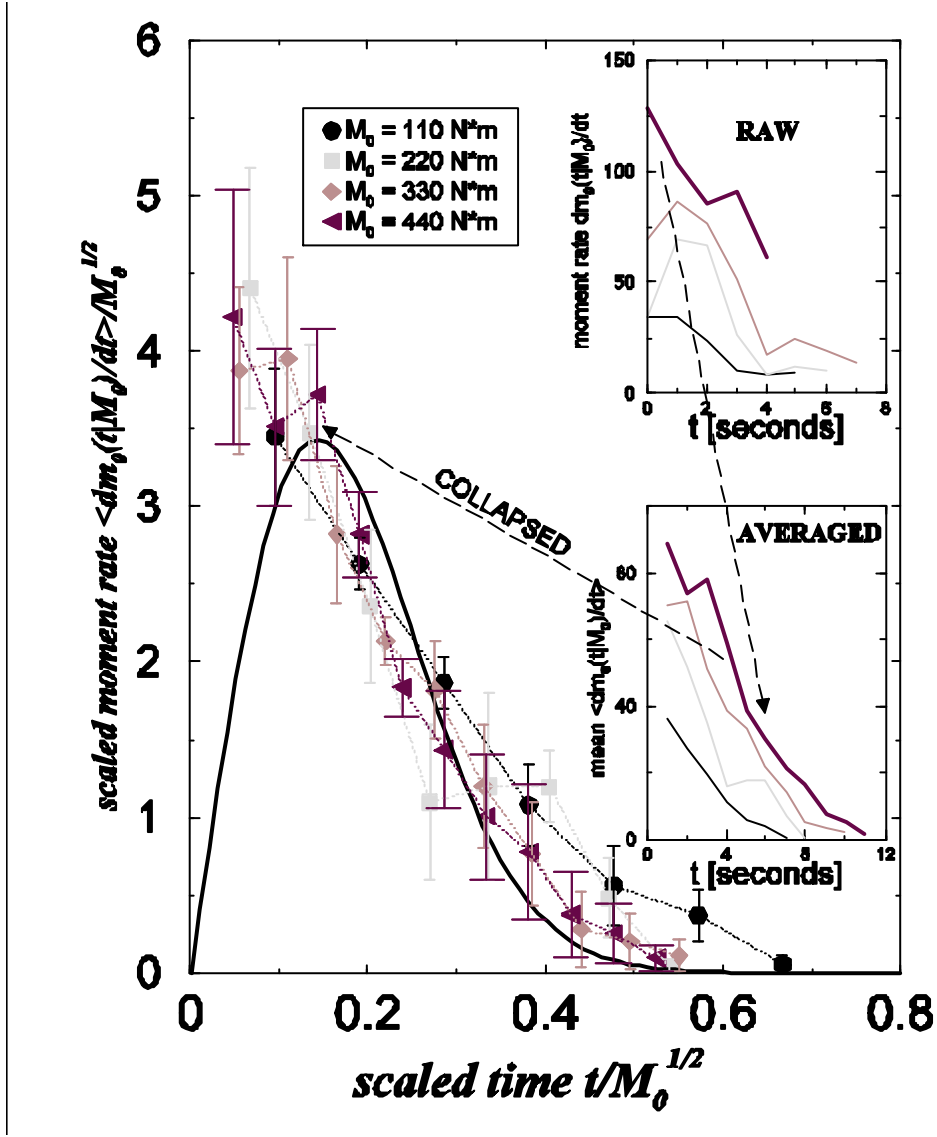

FIGURE 3. A collapse of averaged earthquake pulse shapes, $< dm_0(t|M_0)/dt >$, with the size of the moment $M_0$ in Newton meters within 10% of each size given in the legend respectively. In order to obtain each collapsed moment rate shape, five to ten earthquakes were averaged for each value of $M_0$. The collapse was obtained using the mean field scaling relation[34]: $< dm_0(t|M_0)/dt >/M_0^{1/2} \sim f(t/M_0^{1/2})$ (see text Eq. (6)). In our mean field theory the universal scaling function is $f_{mf}(x) = Axe^{-Bx^2/2}$ where $x = t/M_0^{1/2}$. We plot this functional form (bold curve) with $A = 4$ and $B = 4.9$. Insets: We show the raw data and the averaged data (before collapsed).

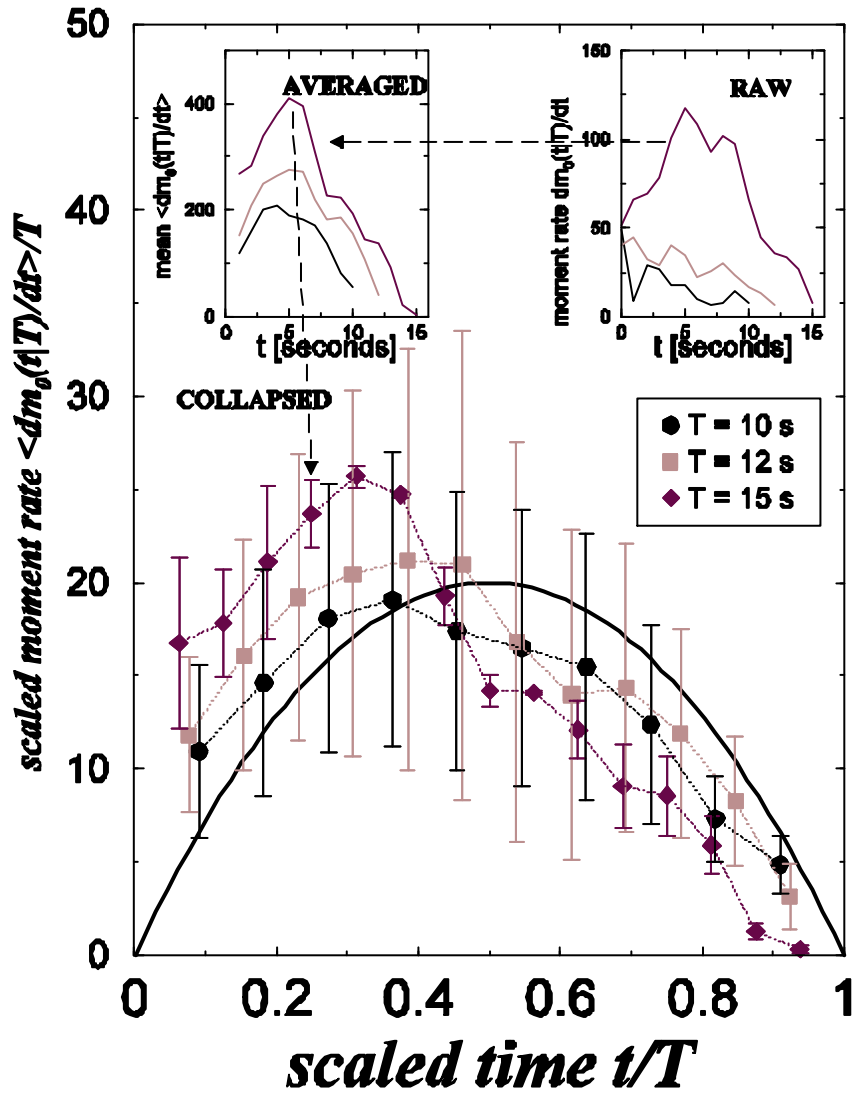

FIGURE 4. A collapse of averaged earthquake pulse shapes, $<dm_0(t\,|\,T)/dt>$ with a duration of T (seconds) within 10% (given in legend), is shown. The collapse was obtained using the mean field scaling relation[33]: $<dm_0(t\,|\,T)/dt>/T \sim g(t/T)$. In order to obtain each collapsed pulse shape, two to ten earthquakes were averaged for each value of *T*. In our mean field theory the universal scaling function is $g_{mf}(x) = Ax(1-x)$ with $x = t/T$. We plot this functional form (bold curve) with A = 80. Note the apparent asymmetry to the left in the observed data while the theoretical curve is symmetric around the maximum. Insets: We show the raw data and the averaged data (before collapsed).

References:

1. B. Gutenberg, C. F. Richter, *Seismicity of Earth and Associated Phenomena* (Princeton Univ. Press, Princeton, 1954).

2. Y. Ben-Zion, *International Handbook of Earthquake and Engineering Seismology* (Academic Press, 2003), Part B, pp. 1857-1875.

3. T. Utsu, *International Handbook of Earthquake and Engineering Seismology*, W. HK Lee, H. Kanamori, P. C. Jennings, C. Kisslinger, Eds. (2002) Part A, pp. 719-732.

4. T. C. Hanks, H. Kanamori, *J. Geophys. Res.*, **84**, 2348 – 2350 (1979).

5. C. Frohlich, S. D. Davis, *JGR*, **98**, 631-644 (1993).

6. Y. Utsu, Y. Ogata, R. S. Matsu'uara, *J. Phys. Earth*, **43,** 1-33 (1995).

7. C. Kisslinger, L. M. Jones, *J. Geophys. Res.*, **96**, 11,947-11,958 (1991).

8. D. S. Fisher, K. Dahmen, S. Ramanathan, Y. Ben-Zion, *Phys. Rev. Lett.* **78**, 4885-4888 (1997).

9. S. G. Wesnousky, *Bull. Seismol. Am.* **84**, 1940-1959 (1994).

10. M. W. Stirling, S. G. Wesnousky, K. Shimazaki, *Geophys. J. Int.*, **124**, 833-868 (1996).

11. S. Zapperi, P. Ray, H. E. Stanley, A. Vespignani, *Phys. Rev. E*, **59**, 5049-5057, (1999).

12. P. Bak, C. Tang, K. Wiesenfeld, *Phys. Rev. A*, **38**, 364-374, (1988).

13. D. Sornette, C. G. Sammis, *J. Phys. I France*, **5**, 607-619 (1995).

14. Rundle, J. B., D. L. Turcotte, R. Shcherbakov, W. Klein, and C. Sammis, *Rev. Geophys.,* **41**(4), 1019, 2003.

15. M. Schroeder, *Fractals, chaos, power laws* (W. H. Freeman and Co., 1991).

16. D. Sornette, *Critical phenomena in natural sciences, Chaos fractals, self organization and disorder: concepts and tools* (Springer-Verlag, 2000).

17. Y. Ben-Zion, J. R. Rice, *J. Geophys. Res.*, **98**, 14109-14131 (1993).

18. Y. Ben-Zion, J. R. Rice, *J. Geophys. Res.*, **100**, 12959-12983 (1995).

19. Y. Ben-Zion, *J. Geophys. Res.*, **101**, 5677-5706 (1996).

20. K. Dahmen, D. Ertas, Y. Ben-Zion, *Phys. Rev. E*, **58**, 1494-1501 (1998).

21. D. Vere-Jones, *Pure and Appl. Geophys.*, **114**, no.4, p. 711-726 (1976).

22. A. P. Mehta, A. C. Mills, K. Dahmen, J. P. Sethna, *Phy. Rev. E.*, **65**, 46139/1-6 (2002).

23. E. T. Lu, R. J. Hamilton, J. M. McTiernan, K. R. Bromond, *Astrophys. J.* **412**, 841 (1993).

24. J. P. Bouchaud, M. Potters, *Theory of Financial Risks: From Statistical Physics to RiskManagement* (Cambridge University, Cambridge, 2000).

25. Y. Ben-Zion, C. G. Sammis, *Pure appl. Geophys.* **160**, 677-715 (2003).

26. J. Liu, K. Sieh, E. Hauksson, *Bull. Seism. Soc. Am.,* **93**, 1333-1344, (2003).

27. S. Zapperi, P. Cizeau, G. Durin, H. E. Stanley, *Phys. Rev. B.* **58 (10)**, 6353 6366 (1998).

28. G. Durin, S. Zapperi, *Alamos Nat'l Laboratory Archive*, http://xxx.lanl.gov/abs/cond-mat/0106113, (2001).

29. J. P. Sethna, K. A. Dahmen, C. R. Myers, *Nature,* **410**, 242-250 (2001).

30. K. G. Wilson, *Scientific American*, **241**, 158-179, (1979).

31. J. J. Binney, N. J. Dowrick, A. J. Fisher, M. E. J. Newman, *The theory of critical phenomena* (Oxford University Press, 1993).

32. S. L. Bilek, PhD thesis, (University of California, Santa Cruz, 2001), pp. 180.

33. M. C. Kuntz, J. P. Sethna, *Phys. Rev. B.* **62**, 11699-11708 (2000).

34. D. S. Fisher, *Phys. Reports*, **301**, 113-150 (1998).

35. P. M. Mai, P. Spudich, J. Boatwright, *Seis. Res. Lett.*, **74**, 208 (2003).